\documentclass[usegraphicx,useAMS,usenatbib]{mn2e}
\usepackage{times}
\usepackage{amssymb}
\usepackage{amsmath}
\usepackage{graphicx}
\usepackage{euscript}
\usepackage{rotating}
\usepackage{epsfig}

\def\mhz{{\rm\thinspace MHz}}
\def\ghz{{\rm\thinspace GHz}}
\def\cmsqps{{\rm\thinspace cm^{2}~s^{-1}}}

\def\myr{{\rm\thinspace Myr}}

\begin{document}

\newcommand{\Mpc}{\rm\thinspace Mpc}
\newcommand{\kpc}{\rm\thinspace kpc}
\newcommand{\pc}{\rm\thinspace pc}
\newcommand{\km}{\rm\thinspace km}
\newcommand{\m}{\rm\thinspace m}
\newcommand{\cm}{\rm\thinspace cm}
\newcommand{\cmps}{\hbox{$\cm\s^{-1}\,$}}
\newcommand{\cmpssq}{\hbox{$\cm\s^{-2}\,$}}
\newcommand{\cmsq}{\hbox{$\cm^2\,$}}
\newcommand{\cmcu}{\hbox{$\cm^3\,$}}
\newcommand{\pcmcu}{\hbox{$\cm^{-3}\,$}}
\newcommand{\pcmcuK}{\hbox{$\cm^{-3}\K\,$}}

\newcommand{\yr}{\rm\thinspace yr}
\newcommand{\gyr}{\rm\thinspace Gyr}
\newcommand{\s}{\rm\thinspace s}
\newcommand{\ks}{\rm\thinspace ks}

\newcommand{\GHz}{\rm\thinspace GHz}
\newcommand{\MHz}{\rm\thinspace MHz}
\newcommand{\Hz}{\rm\thinspace Hz}

\newcommand{\K}{\rm\thinspace K}

\newcommand{\Kpcmc}{\hbox{$\K\cm^{-3}\,$}}

\newcommand{\g}{\rm\thinspace g}
\newcommand{\gpcm}{\hbox{$\g\cm^{-3}\,$}}
\newcommand{\gpcmps}{\hbox{$\g\cm^{-3}\s^{-1}\,$}}
\newcommand{\gps}{\hbox{$\g\s^{-1}\,$}}
\newcommand{\Msun}{\hbox{$\rm\thinspace M_{\odot}$}}
\newcommand{\Msunpc}{\hbox{$\Msun\pc^{-3}\,$}}
\newcommand{\Msunpkpc}{\hbox{$\Msun\kpc^{-1}\,$}}
\newcommand{\Msunppc}{\hbox{$\Msun\pc^{-3}\,$}}
\newcommand{\Msunppcpyr}{\hbox{$\Msun\pc^{-3}\yr^{-1}\,$}}
\newcommand{\Msunpyr}{\hbox{$\Msun\yr^{-1}\,$}}

\newcommand{\MeV}{\rm\thinspace MeV}
\newcommand{\keV}{\rm\thinspace keV}
\newcommand{\eV}{\rm\thinspace eV}
\newcommand{\erg}{\rm\thinspace erg}
\newcommand{\Jy}{\rm Jy}
\newcommand{\ergpcmc}{\hbox{$\erg\cm^{-3}\,$}}
\newcommand{\ergcmcups}{\hbox{$\erg\cm^3\ps\,$}}
\newcommand{\ergpcmps}{\hbox{$\erg\cm^{-3}\s^{-1}\,$}}
\newcommand{\ergpcmsqps}{\hbox{$\erg\cm^{-2}\s^{-1}\,$}}
\newcommand{\ergpcmsqpspA}{\hbox{$\erg\cm^{-2}\s^{-1}$\AA$^{-1}\,$}}
\newcommand{\ergpcmsqpspsr}{\hbox{$\erg\cm^{-2}\s^{-1}\sr^{-1}\,$}}
\newcommand{\ergpcmcups}{\hbox{$\erg\cm^{-3}\s^{-1}\,$}}
\newcommand{\ergps}{\hbox{$\erg\s^{-1}\,$}}
\newcommand{\ergpspmp}{\hbox{$\erg\s^{-1}\Mpc^{-3}\,$}}
\newcommand{\keVpcmsqpspsr}{\hbox{$\keV\cm^{-2}\s^{-1}\sr^{-1}\,$}}

\newcommand{\dyn}{\rm\thinspace dyn}
\newcommand{\dynpcmsq}{\hbox{$\dyn\cm^{-2}\,$}}

\newcommand{\kmps}{\hbox{$\km\s^{-1}\,$}}
\newcommand{\kmpspmp}{\hbox{$\km\s^{-1}\Mpc{-1}\,$}}
\newcommand{\kmpspMpc}{\hbox{$\kmps\Mpc^{-1}$}}

\newcommand{\Lsun}{\hbox{$\rm\thinspace L_{\odot}$}}
\newcommand{\Lsunppc}{\hbox{$\Lsun\pc^{-3}\,$}}

\newcommand{\Zsun}{\hbox{$\rm\thinspace Z_{\odot}$}}
\newcommand{\gauss}{\rm\thinspace gauss}
\newcommand{\arcsecond}{\rm\thinspace arcsec}
\newcommand{\chisq}{\hbox{$\chi^2$}}
\newcommand{\delchi}{\hbox{$\Delta\chi$}}
\newcommand{\ph}{\rm\thinspace ph}
\newcommand{\sr}{\rm\thinspace sr}

\newcommand{\pccm}{\hbox{$\cm^{-3}\,$}}
\newcommand{\psqcm}{\hbox{$\cm^{-2}\,$}}
\newcommand{\pcmsq}{\hbox{$\cm^{-2}\,$}}
\newcommand{\pmpc}{\hbox{$\Mpc^{-1}\,$}}
\newcommand{\pmpccu}{\hbox{$\Mpc^{-3}\,$}}
\newcommand{\ps}{\hbox{$\s^{-1}\,$}}
\newcommand{\pHz}{\hbox{$\Hz^{-1}\,$}}
\newcommand{\pcmK}{\hbox{$\cm^{-3}\K$}}
\newcommand{\phpcmsqps}{\hbox{$\ph\cm^{-2}\s^{-1}\,$}}
\newcommand{\psr}{\hbox{$\sr^{-1}\,$}}
\newcommand{\pspsqas}{\hbox{$\s^{-1}\,\arcsecond^{-2}\,$}}

\newcommand{\ergpspcmpK}{\hbox{$\erg\s^{-1}\cm^{-1}\K^{-1}\,$}}

\title{Precession of the Super-Massive Black Hole in NGC 1275 (3C 84)?}
\author[Dunn, Fabian \& Sanders]
{\parbox[]{6.in} {R.J.H. Dunn\thanks{E-mail: rjhd2@ast.cam.ac.uk},
    A.C. Fabian and J.S. Sanders\\
    \footnotesize
    Institute of Astronomy, Madingley Road, Cambridge CB3 0HA\\
  }}

\maketitle

\begin{abstract}
The X-ray holes at the centre of the Perseus Cluster of galaxies are
not all at the same position angle with respect to the centre of the cluster.
This configuration would result if the jet inflating the bubbles is
precessing, or moving around, and the bubbles detach at different times.  The orientations which best fit the observed travel
directions are an inclination of the precession axis to the
line of sight of $120^{\circ}$ and an opening angle of $50^{\circ}$.
From the timescales for the bubbles seen in the cluster, the
precession timescale, $\tau_{\rm prec}$, is around $3.3 \times 10^7
\yr$.  The bubbles rising up through different parts of the cluster may
have interacted with the central cool gas, forming the whorl of
cool gas observed in the temperature structure of the cluster.  The
dynamics of bubbles rising in fluids is discussed.  The
conditions present in the cluster are such that oscillatory motion,
observed for bubbles rising in fluids on Earth, should take
place. However the timescale for this motion is longer than that taken for the bubbles to
evolve into spherical cap bubbles, which do not undergo a path
instability, so such motion is not expected to occur.
\end{abstract}


\section{Introduction}

High resolution radio observations of a number of extragalactic radio
sources show that the opposed jets contain inversion symmetries, which provides
evidence that the central engine is precessing, e.g. NGC 326
\citep{Ekers1978} and NGC 315 \citep{Bridle1979}.  In a review of
quasars and radio galaxies which exhibit bent jets, of which some are
inversion-symmetric about the core, \citet{Gower_1982} present ten
sources whose morphology can be explained using a precessing jet
with periods ranging between $10^4\to10^7\yr$.  The extraordinary
galactic object SS433 \citep{HjellmingJohnston81} is successfully described with a
precessing jet model by \citet{Margon84}, and has been the subject of a
recent study by \citet{Blundell}.

The radio source at the centre of the Perseus Cluster, 3C84, is an
extended source and shows an ``{\sf S}'' or ``{\sf Z}''-shaped morphology
at $\ghz$ frequencies.  These lobes have been observed to
anti-correlate spectacularly with the observed X-ray emission from the
cluster \citep{Bohringer,ACF_complex_PER00}.  These depressions in the X-ray emission are thought to be the result of the jet from
the central engine blowing
bubbles of relativistic plasma into the thermal Intra-Cluster Medium,
pushing the ICM aside \citep{GullNorthover,Bohringer}.  When their buoyancy velocity
is greater than their expansion velocity, the bubbles are
expected to detach and rise up buoyantly through the ICM \citep{Churazov00}.  This is
observed in the Perseus Cluster, to the North-West and to the South of
the core where there are depressions in the X-ray emission which do not have
any associated $\ghz$ radio emission; however, at lower frequencies
(e.g. $330\mhz$) ``spurs'' of
emission are observed pointing towards these so-called ``Ghost
bubbles'' \citep{Celotti02}.

\begin{figure*}
\includegraphics[width=0.95 \columnwidth]{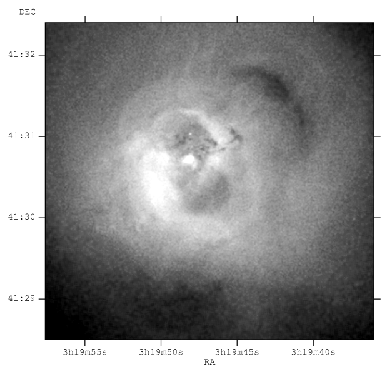}
\includegraphics[width=0.95 \columnwidth]{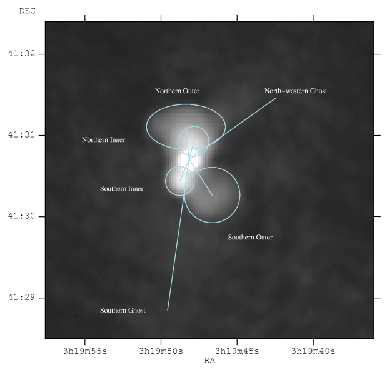}
\caption{\label{Xray_radio_image} {\scshape left:} The X-ray emission from the centre of the
  Perseus Cluster from the 200 ks observation of
  \citet{ACF_deep_PER03}, corrected and accumulatively smoothed with a
  $\sigma=15$. {\scshape right:} A $330 \mhz$ radio map of
  3C84.  The shapes used for the
  different bubbles and the lines joining their centres to the radio
  nucleus are shown along with their names.}
\end{figure*}

The two pairs of bubbles which are clearly visible in the X-ray
emission from the cluster belong to two different ``generations,''
the young, radio-active pair and the older ``ghost'' pair.  The
bubbles are observed at different position angles with respect to the
radio core in the centre of the cluster.  In Section \ref{Radioemission} we discuss
the morphology of the radio emission at the centre of the cluster,
linking in other features in Section \ref{otherfeatures}.  Section \ref{model} outlines
the model for the precessing jet, the resultant parameters of
which are presented in Section \ref{parameters}.  We calculate the timescales
for the precession in Section \ref{timescales} and discuss the model in Section
\ref{discussion}.  We put forward two candidates for causing the precession in
Section \ref{causeprec} and discuss the dynamics of rising bubbles in Section \ref{daVinci}.

\section{The Radio Bubbles of the Perseus Cluster}
\label{Radioemission}

The extended radio emission from the radio source does not
decrease smoothly from the core; there is a sharp drop
in radio flux in the southern lobe (see Fig. \ref{Xray_radio_image} \& \citealt{Pedlar90}).  This could be because there are
two bubbles present, an older one which is in the process of rising buoyantly
and a younger one which is still attached to the jet and being
inflated.  The X-ray emission also
shows features in the region where the drop in radio flux occurs, there is a swath of
excess X-ray emission which could arise from the ICM flowing in behind
the recently detached bubble (Fig. \ref{Xray_radio_image}).  The shape of
the X-ray decrement corresponding with this ``Southern Outer'' bubble
is also similar to that of a spherical cap bubble.

\begin{figure}
\includegraphics[width=0.95 \columnwidth]{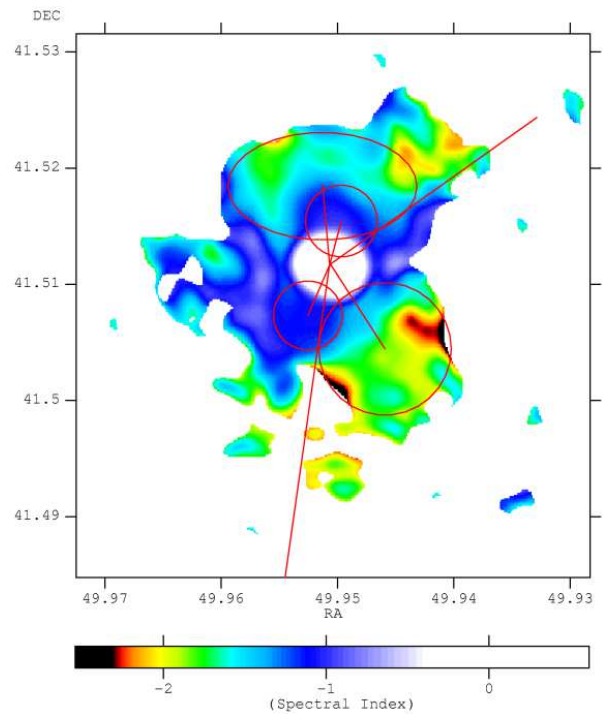}
\caption{\label{spectInd} The 330 to 1500 $\mhz$ spectral index map of
the Perseus Cluster (G. Taylor, private comm.) showing the shapes for
the bubbles.}
\end{figure}

There is no equivalent sharp drop in the radio flux in the northern lobe, which,
if there are also two bubbles present in this lobe, may result if
younger one is behind the older one.  This could soften the change in flux
between the two bubbles as projected on the sky.  In the case that the
northern and southern radio lobes contain two bubbles each, we 
identify them as shown in Fig. \ref{Xray_radio_image}.  A spectral
index map of the central regions of Perseus (Fig. \ref{spectInd}) shows that the bubbles we
have labelled as Outer have a steeper index than the Inner ones ($-1.7$
vs. $\sim -1.2$ for $S_\nu\propto\nu^\alpha$).  This supports the idea that these Outer bubbles are
older than the Inner ones, as the electrons have aged, steepening the
spectrum.

What can be seen from these observations is that the bubbles do not
necessarily travel in the same directions when they detach.  This is
most obvious from the ghost bubbles.  This may be due to the bubbles
being somehow deflected from their initial path.  Another explanation
may be that the jet produced by the Super-Massive Black Hole (SMBH) is
precessing, and the bubbles detach at different phases in the
precession cycle.  This analysis assumes that
the bubbles travel outward from the centre in the direction the jet
was pointing in at the time that they detached, with the precession time
 longer than the bubble inflation time.  The current positions
of the radio bubbles allow some constraints to be placed on the
orientation and timescales of the precession.  

However, Leonardo da Vinci observed that bubbles rising in liquids can move in a zig-zag
fashion \citep{LeonardodV}, and this may also be an explanation for the difference in
observed position angles of the bubbles.  However the analysis in
Section \ref{daVinci} leads to the conclusion that this behaviour is not
relevant in the Perseus Cluster.

The northern ghost bubble has been likened to a spherical cap bubble
rising through fluid \citep{ACF_deep_PER03}.  The shape of the bubble
is very similar to such a bubble in profile, and so can be reasonably
expected to be travelling out from the centre of the cluster in the
plane of the sky.  Observations of the
H$\alpha$ filaments which are found around NGC1275 match some of the
flow lines expected behind such a bubble very
closely. \citet{Hatch2005inprep} show that the line of sight
velocities for the Horseshoe-filament which sits behind the
North-Western Ghost bubble are very small for the sections of the
filament that point towards the centre of the cluster, implying
that they probably move at $\sim 90^{\circ}$
to the line of sight.

Very Long Baseline Interferometry (VLBI) maps of the region close to
the core show an asymmetry between the southern and northern jets.
While the southern jet is seen at all radio frequencies, the inner
parts of the northern jet are not detected below $15\ghz$.  At the
lower frequencies free-free absorption has been shown to be a reason
why not all of the jet is seen \citep{Walker2000}.  At higher frequencies, the
inclination of the jet would play a role, with the southern one being
inclined towards us.  The amount of beaming, and hence asymmetry
depends on the jet's bulk velocity.  This inclination of the jet
matches the orientation of the larger scale lobes; the northern Inner
on may be behind the older Outer one.

Measurements of the VLBI jets can give some constraints on the current
jet direction.  \citet{Krichbaum92_Per} calculate that the jet is at $\le 14.4^{\rm
  o}$ to the line of sight, and even $\le 2.7^{\circ}$ close to the
core for the milliarcsecond jet; however they find the jet
to be at $39.4^{\circ}\le \alpha \le 58.2^{\circ}$ to the line of
sight when observed at arcsecond scales, requiring it to bend by $25 -
45 ^{\circ}$.  We take the jet to be at $\sim 10^{\circ}$ to the line
of sight.

\section{Other Features}\label{otherfeatures}

\subsection{Low Temperature Whorl and Ancient Bubbles}\label{AncientWhorl}

\begin{figure}
\includegraphics[width=0.95 \columnwidth]{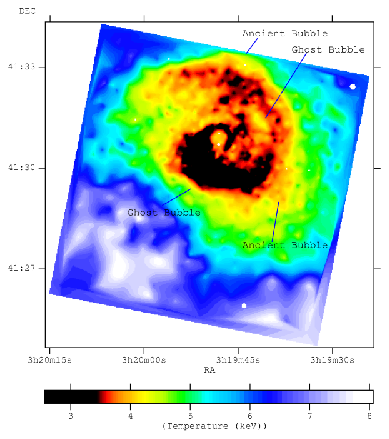}
\caption{\label{Tempwhorl} A smoothed version of the temperature map of the Perseus
  Cluster \citep{Sanders05}, interpolated with the \textsc{natgrid}
  natural neighbour interpolation library
  (http://ngwww.ucar.edu/ngdoc/ng/ngmath/natgrid/nnhome.html).  For
  more details see \citet{ACF_3C294}.  There
  are some edge effects in the image resulting from the
  smoothing. White circles are excluded point sources, listed in
  \citet{SandersMap04}.  The approximate locations of the older
  bubbles present in the cluster are shown.} 

\end{figure}

A spiral structure in the Perseus Cluster was first noticed by \citet{Churazov00} and is seen in
the temperature maps of \citet{SandersMap04} (Fig. \ref{Tempwhorl}).
\citet{ACF_complex_PER00} suggested that this whorl could arise from the
merger of a cool subcluster, as stripping of gas as it passed
through the ICM would have left a wake.  It could also form as the result of cooling gas falling into
the centre of the cluster if the gas started with some initial angular
momentum.  However, bulk motions of the gas would be expected to have
an effect on the H$\alpha$ filaments (Section \ref{halpha}) and as they are
linear this is unlikely as an explanation.
If the jet is precessing, then the bubbles rise through the ICM at
different position angles.  They would push low temperature material in front of
them or drag it up in their wake as they rise.  The North-Western Ghost bubble
falls on the inner edge of this whorl.  There is also a hint that
there is an even older (``ancient'') bubble to the
North, seen as a crescent-shaped bump on the outer surface of this whorl (3h19m45s,
41:33$^{\circ}$) (Fig. \ref{Tempwhorl}).  If this is an ancient bubble, then the decrement in
the X-ray emission is now so small that it cannot be clearly seen in
the X-ray images of the cluster.  However the change in the
temperature of the ICM caused by its motion up through the cluster is
still visible. 

This whorl is most easily explained with a precession
model where the precession axis is face-on to us with a large ($\sim
90 ^{\circ}$) opening angle.  As the bubbles detach
from such a slowly clockwise-precessing jet they push
up cooled, low temperature gas from the centre of the cluster.  The
next bubble lifts material from a slightly different position angle.
This low temperature gas could be connected together by magnetic
fields present in the cluster.  Thus the portions of gas which are not
lifted/pushed up by the bubbles are pulled up by the portions of gas
that are, and so a whorl forms as opposed to discrete portions of gas
around the rising bubbles.

The interaction of the Northern bubbles (Ancient and Ghost) with the
whorl is fairly clear, the Southern Outer bubble also sits behind it
and so could just be starting to lift this section of the
low-temperature gas out of the centre of the cluster.  There is some
extension of this cool gas to the West in the same direction as a high
abundance ridge presented in \citet{Sanders05}.  This has also been interpreted as the
remnants of an ancient bubble which lifted up the enriched material from
the centre of the cluster.  The interaction of the Southern Ghost bubble with the low temperature
gas is not as clear, but it sits just South the coolest gas at the
centre of the cluster.  It may have lifted cooler gas out from the
centre, but if it were travelling close to the line of sight, in projection the
distance the gas would have appeared to have been moved is not that far.


It would be expected that if this whorl is the result of the
jet/bubbles interacting with the ICM, then two spirals should be seen,
one resulting from each jet.  However, it seems as if there is only
one.  The explanation for this could be that the cooling has
resulted in an asymmetric arrangement of gas.  The northern
bubbles could then have broken through this gas, whereas the southern
ones have detached in such an orientation that the spiral appears
continuous.

The whorl may, of course, be the result of a chance projection of two
rings; a larger outer one which forms the northern and south-western
parts of the whorl and an inner one around the northern outer bubble
and the eastern part of the southern younger bubbles.  We do not offer
an explanation of the formation of this structure but merely put
forward than the the whorl may not be one single structure.

\subsection{H$\alpha$ Filaments}\label{halpha}

The large H$\alpha$ nebula around NGC
1275 has been studied at a variety of wavelengths \citep{Minkowski57, Lynds70}, and deep
images obtained by the WIYN
telescope show extended filaments which stretch over $50\kpc$ from the
central galaxy \citep{Conselice01}.  \citet{ACF_Halpha_PER03} showed
that there was a correspondence between the soft X-ray and the H$\alpha$
filaments.  They also showed that the filaments may represent the flow
lines of the ICM in the wake of passing bubbles, for example the
``horseshoe'' filament which sits behind the North-Western ghost
bubble.  There is a large collection of filaments stretching up
northwards from the cluster in the direction of the possible ancient
bubble mentioned above.  If the northern ancient bubble is a real feature and the
filaments are linked with it, then they are likely to have been drawn
up out of the cluster by the passage of the bubble.  \citet{Sanders05}
also find evidence for filaments pointing in the
direction of the South-Western ancient bubble.

\subsection{Spiral Arms in NGC1275}

In Hubble Space Telescope (HST) images of the nucleus of NGC 1275,
ripples were noticed in the galaxy light \citep{Holtzman92, Carlson98}
which resemble the arms of spiral galaxy.  These ripples spiral in
the same sense as the low-temperature whorl seen in the X-ray gas
(Fig.  \ref{Tempwhorl}) and so may be connected.  If there is a
precessing black hole at the centre of the galaxy, then the precession
of the jet may be linked with these features.

\section{Precessing Jet Model}\label{model}

From the shape of the low temperature whorl, the precessing jet
may arranged such that the precession axis points close to
the line of sight, with an opening angle of around $90^{\circ}$ (wide cone), or,
as there is some reflection symmetry about the East-West axis a narrow
cone precession where the precession axis is aligned North-South with
a smaller opening angle.

\begin{figure}
\includegraphics[width=0.95 \columnwidth]{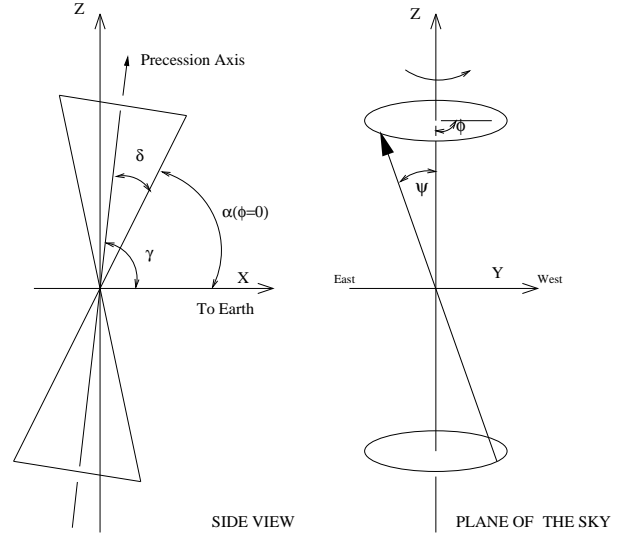}
\caption{\label{Model} The model of the precessing jets.}
\end{figure}

To describe the precessing jet, the model outlined in
\citet{Barker1981} has been adapted and is shown in Fig. 
\ref{Model}, where $\gamma$ is the angle the
precession axis makes with the line of sight, and $\delta$ is the
angle the jet makes with the precession axis.  The angle the jet makes
with the line of sight, $\alpha$, is given by:
\begin{equation}\label{cosalpha}
\cos\alpha=\cos\gamma \cos\delta + \sin\gamma \sin\delta \cos\phi,
\end{equation}
where $\phi$ is the phase angle of a clockwise rotating jet, and is
$0^{\circ}$ when the jet points towards Earth.  When $\alpha=90^{\rm
  o}$ then \mbox{$\cos\phi=-\cot\gamma \cot\delta$.}

If $\psi$ is the angle between projected direction of the jet on the
plane of the sky and the projected direction of the precession axis, then
\begin{equation}\label{tanpsi}
\tan\psi=\frac{\sin\delta\sin\phi}{\sin\gamma\cos\delta-\cos\gamma\sin\delta\cos\phi}.
\end{equation}
An extreme value of $\tan\psi$ occurs at
$\cos\phi=\tan\delta\cot\gamma$.

\begin{table}  \caption{\scshape \label{SkyAngles_Times} Bubble
    directions and Timescales.}
\begin{tabular}{lllll}
&\multicolumn{2}{c}{Angle from North $(0^{\rm
    o})$$^\dagger$}&\multicolumn{2}{c}{Timescales$^\ddagger$/$10^7\yr$}\\
Lobe & N & S & N & S\\
\hline
Jet     & $0$&$180$&$-$&$-$\\
Inner   & $15$&$205$&$1.1$&$1.0$\\
Outer   & $355$&$145$&$1.5$&$1.9$\\
Ghost   & $55$&$190$&$7.5$&$7.1$\\
Ancient & $15$&$130$&$10.0$&$10.0$\\
\end{tabular}
\begin{quote}
\noindent$^\dagger$ The angles are measured Clockwise when looking at the figures
presented here.\\
$^\ddagger$ The timescales (ages) have not been adjusted for projection effects.\\
{\scshape Note:} All angles and timescales are approximate.  The
Ancient bubbles are discussed in Section \ref{AncientWhorl}.
\end{quote}
\end{table}

The angles from North (clockwise as seen in the Figures) and the timescales for all the bubbles are
listed in Table \ref{SkyAngles_Times}.  We have used the sound speed timescale for the Inner and Outer
bubbles and the buoyancy timescale for the Ghost and Ancient
bubbles.  The buoyancy timescales estimate the time taken for a
bubble of the currently observed size to rise buoyantly from the centre of the
cluster through a uniform medium of that it is currently
observed to be in.  The sound speed timescale estimates the time taken
for a bubble of the currently observed size to be blown if the
expansion is subsonic.  These two timescale measures are different,
but provide the best estimate of the age of the bubbles at different
stages in their formation (for further discussion of these timescales see
\citet{DunnFabian04,DunnFabian05}).  The difference in timescales
between the Inner and Outer bubbles is small
($0.4\to0.9\times 10^7 \yr$), whereas the differences are larger
between the Outer/Ghost and Ghost/Ancient bubbles.  This may be a result of the different estimates of the timescales that
have been used.  The buoyancy timescale of a bubble, being a rise time, is generally
larger than the sound speed timescale, which is an expansion time, and so the jump in ages between
the Outer and Ghost bubbles may be a result of this.  It is unlikely
that the differences are largly the result of anisotropies in the ICM,
as the pressure exerted on the bubbles changes only by a factor of
$\sim 2$ \citep{SandersMap04}.  Projection
effects may remove (or enhance) this jump as the sound speed timescale
is unaffected by the orientation, being an expansion timescale;
whereas the buoyancy timescale does depend on the inclination of the
rise-path of the bubble to the line of sight.

\section{Calculation of the Precession Parameters}\label{parameters}

When calculating the best fitting parameters, the angle of the jet to
the line of sight has not
been used as a constraint as the actual angle is uncertain.  However, as the difference
in the emission from the Northern and Southern parts
implies that the southern jet is coming towards us, this has to be
satisfied by the parameters.

As the Northern Ghost bubble appears to be moving close to the plane
of the sky we assume that $\alpha=90^{\circ}$ for this bubble.  The shape of the
Southern Outer bubble also appears to be such that it is also moving
close to the plane of the sky; this has also been used to constrain
the possible angles.  However, as there are only two bubbles which
provide constraints on $\alpha$,
and these appear to move in the
plane; a close to face-on precession axis with a high value for
$\delta$ (around $90^{\circ}$) would
also fit the data. 

Although this arrangement is a possiblity, such an extreme opening angle
would be very unusual for a black hole of this mass.  Also it would be expected that
the bubbles would have detached at all position angles over the course
of a few precession timescales.  However the position angles of the
northern bubbles are very similar to those of the southern bubbles,
almost mirror symmetry about an East-West axis.

From Equations \ref{cosalpha} and \ref{tanpsi} and the observed
projected angles on the sky, the best fit occurs when the projected
angle of the precession axis on the sky is $10^{\circ}$ (clockwise of
North as seen in Fig. \ref{Xray_radio_image}).  Then there are various
combinations of $\gamma$ and $\delta$ which fit the observed data.
However, we choose those with the smallest $\delta$ (precession opening angle)
are $\gamma=60^{\circ}$ or $120^{\circ}$ with $\delta=50^{\circ}$ ($\pm 5^{\circ}$).

To decide which of the two scenarios is the more plausible we look at
the VLBI jet.  The Southern jet is brighter than the Northern jet,
which implies that it is Doppler beamed, and so is thought to be
travelling towards us.  The angles of the jet to the line of sight
close to the radio core are around $10^{\circ}$
\citep{Krichbaum92_Per}.  This implies that the precession axis
is pointing away from Earth in the North; hence $\gamma=120^{\circ}$
is preferred as the orientation.  


We assume that the jets are precessing anti-clockwise when viewed down
the precession axis as shown in Fig. \ref{Model}.
There are usually two possible phase angles ($\phi$) for those bubbles
which have no constraint on $\alpha$.  The single values have been chosen from the distance
travelled by the bubble and also from the observed shape; for example,
the Southern Ghost bubble could have travelled at either $70^{\circ}$ or $170^{\circ}$
to the line of sight.  We choose  $70^{\circ}$ as, if it were
travelling at $170^{\circ}$, its true age would be much older than
calculated and the X-ray depression would not be expected to be visible.  It also
appears to be more in cross-section rather than face on.  Similar arguments
have been used to choose the angles for the other bubbles.

\begin{table}  \caption{\scshape \label{PrecPhase_Times} Phase angles
    and adjusted timescales}
\begin{tabular}{lllllll}
&\multicolumn{2}{c}{Phase angle,$^\dagger$ $\phi$}&\multicolumn{2}{c}{Angle
  to l.o.s.}&\multicolumn{2}{c}{Timescales$^\ddagger$}\\
Lobe & N & S & N & S & N & S\\
\hline
Jet     & $182$ & $182$&$170$ & $10$&$-$ & $-$\\
Inner   & $6/179$ & $19/177$&$70/170$ & $108/10$&$1.1$ & $1.0$\\
Outer   & $341$ & $293$&$72$ & $94$&$1.5$ & $1.9$\\
Ghost   & $67$ & $0$&$94$ & $110$&$7.6$ & $8.1$\\
Ancient & $6$ & $247$&$70$ & $55$&$11.4$ & $15.3$\\
\end{tabular}
\begin{quote}
\noindent
$^\dagger$ The phase angle is measured for the Northern jet, and the
precession is assumed to be clockwise as see in the images presented
here.\\
$^\ddagger$ The timescales have now been adjusted for projection
effects. They are accurate to around $20$ per cent.  The sound speed
expansions timescale has been used for the Inner and Outer bubbles,
the buoyancy rise time for the Ghost and Ancient bubbles.\\
{\scshape Note:} All angles and timescales are approximate. Timescales
in $10^7\yr$.
\end{quote}
\end{table}

\section{Timescales}\label{timescales}

To calculate the precession timescales we use the change in phase between the
Ghost and the Outer bubbles, as these are likely to be moving
rectilinearly and have their ages estimated using the same
method.  In the assumption that the jet is precessing
clockwise, then for the Northern jet
$\Delta\phi=274^{\circ}$ and $\Delta t=6.1\times 10^7 \yr$; for the
Southern jet $\Delta\phi=293^{\circ}$ and $\Delta t=6.2\times 10^7
\yr$.  This leads to precession timescales of $8.0$ and $7.6\times
10^7 \yr$ for the Northern and Southern jets respectively, which agree
very well.
For the other precession timescales see Table \ref{PrecTimes}.

\begin{table}  \caption{\scshape \label{PrecTimes} Precession Timescales}
\begin{tabular}{lr@{ $=$ }lr@{ $=$ }l}
 & \multicolumn{2}{c}{North} &\multicolumn{2}{c}{South} \\
\hline
Ghost$\to$Outer&$\Delta\phi$&$274^{\circ}$&$\Delta\phi$&$293^{\circ}$\\
&$\Delta t$&$6.1$&$\Delta t$&$6.2$\\
&$\tau_{\rm prec}$&$8.01$&$\tau_{\rm prec}$&$7.62$\\
& \multicolumn{4}{c}{Average $\tau_{\rm prec}=7.8$}\\
\hline
Ancient$\to$Ghost&$\Delta\phi$&$61^{\circ}$&$\Delta\phi$&$113^{\circ}$\\
&$\Delta t$&$3.8$&$\Delta t$&$7.2$\\
&$\tau_{\rm prec}$&$22.4$&$\tau_{\rm prec}$&$22.9$\\
& \multicolumn{4}{c}{Average $\tau_{\rm prec}=22.7$}\\
\hline
Outer$\to$Jet&$\Delta\phi$&$201^{\circ}$&$\Delta\phi$&$249^{\circ}$\\
&$\Delta t$&$1.5$&$\Delta t$&$1.9$\\
&$\tau_{\rm prec}$&$2.68$&$\tau_{\rm prec}$&$2.75$\\
& \multicolumn{4}{c}{Average $\tau_{\rm prec}=2.7$}\\
\hline
\end{tabular}
\begin{quote}
\noindent{\scshape Notes:} $\Delta t$ and $\tau_{\rm prec}$ in $\times 10^7\yr$.
\end{quote}
\end{table}

Instead of using the Inner bubbles we used
those for the VLBI jets when comparing to the Outer bubbles.  The difference in ages between the Inner and
Outer bubbles is very short, which would give a very rapid precession
timescale.  In the assumption that the VLBI jet is powering the Inner
bubbles, it is more sensible to choose the parameters for the jets, as
the Inner bubbles may still be forming.  The
change in phase angle between the Inner bubbles $(\phi \sim 178^{\rm
  o})$ and the jets $(\phi \sim 182^{\circ})$ is in the correct sense for a
counter-clockwise precessing jet (in the assumption that the Inner
bubbles are $\phi \sim 178^{\circ}$).

There is good agreement between the pairs of bubbles on the value of the
precession timescale, $\tau_{\rm prec}$.  There is a large discrepancy
between the different bubble pairs.  The precession may be speeding
up, which would explain the decrease in the precession time; however
the acceleration would be very rapid - over the course of $\sim1.5$
rotations ($\sim 600^{\circ}$) the precession timescale changes by
around a factor of $10$, from $0.2 \gyr$ to $0.02\gyr$.

It would be expected that the precession timescale would
be the same over many cycles and so we assume that, as the precession
timescale for the Outer$\to$Jet is the shortest, then this is the
precession time.  The $\tau_{\rm prec}$ for the Ghost$\to$Outer and
the Ancient$\to$Ghost are much longer, however if an extra precession
cycle occurred between the Ghost and Outer bubbles;
i.e. $\Delta\phi=(274+360)=634^{\circ}$ for the Northern pair, this
would give a precession
timescale of $3.46\times10^7 \yr$ which is much closer to that for the
Outer$\to$Jet.  For the
number of cycles required for the other bubbles see Table \ref{CycleNo} .

\begin{table}  \caption{\scshape \label{CycleNo} Precession Cycles}
\begin{tabular}{lr@{ $=$ }lr@{ $=$ }l}
 & \multicolumn{2}{c}{North} &\multicolumn{2}{c}{South} \\
\hline
Ghost$\to$Outer&$n$&$1$&$n$&$1$\\
&$\tau_{\rm prec}$&$3.46$&$\tau_{\rm prec}$&$3.42$\\
\hline
Ancient$\to$Ghost&$n$&$1$&$n$&$2$\\
&$\tau_{\rm prec}$&$3.25$&$\tau_{\rm prec}$&$3.11$\\
\hline
Outer$\to$Jet&$n$&$0$&$n$&$0$\\
&$\tau_{\rm prec}$&$2.68$&$\tau_{\rm prec}$&$2.75$\\
\hline
\end{tabular}
\begin{quote}
\noindent{\scshape Notes:} $\Delta t$ and $\tau_{\rm prec}$ in $\times 10^7\yr$.  $n$ from $\tau_{\rm
  prec}=\frac{360\Delta t}{\Delta\phi +n360}$
\end{quote}
\end{table}

Table \ref{CycleNo} shows that, as a result of the large difference in
timescales for the Ancient Northern and Southern bubbles, an extra
precession cycle is required if the precession timescale is fairly
constant.  This could result if we have mis-identified the generation
the southern Ancient bubble belongs to; i.e. that it is from an even
older set, and the features of the true Ancient bubble have been
lost.  It may also mean that the bubbles detach at vastly different
times; the timescales of these old bubbles are calculated using
buoyancy arguements, so if the northern Ancient bubble detached a long
time after the southern one, or rose much more slowly, then this may be
the result.

The $\tau_{\rm  prec}$ for the Outer$\to$Jet are smaller than the other
estimates but this can be explained as the bubbles, which correspond
to the current position of the jets, are still forming and so the
precession timescale for these two may easily be larger.  However
these calculations show that the precession timescale for the SMBH in
the centre of the Perseus Cluster is around $3.3 \times 10^7 \yr$.  

\section{Discussion}\label{discussion}

In the calculation of the precession parameters we have assumed that
the directions in which the bubbles are observed to be moving match
the directions of the jets at the time when the bubbles were just
forming.  As the timescales for the formation of a bubble ($t_{\rm
  sound}=1\to2\times 10^7\yr$) are shorter than the precession timescale
for the jet ($\sim3.3 \times 10^7 \yr$), then the bubble will finish forming
before the jet has moved on (though this would depend on the
exact formation mechanisms of the bubbles).  The travel direction of
the bubbles may therefore indicate the average phase angle of the jet during
the time of their formation.   However, some bubbles may rise up in a
direction corresponding to the direction of the jet towards the end of
their inflation phase, and others in the direction from the beginning
of their formation.  There is therefore some uncertainty in the phase
of the precession corresponding to the bubbles and hence in the
inferred precession times.  

We have also assumed that the bubbles are not deflected from their
path once they start to rise up.  The bubbles would be expected to
travel up out of the potential well of the cluster, however a denser
region of the ICM may deflect the bubbles to different observed
position angles.

The ``ancient'' bubbles that have been used to calculate the
timescales may not be the remnants of very old bubbles, but just
chance features in the ICM which have been misinterpreted.  However
the fact that the Northern Ancient bubble sits close to the
low-temperature whorl, and has H$\alpha$ filaments pointing from the
centre of the cluster in its direction; the high abundance ridge which
corresponds to an extension of the radio mini-halo and the
low-temperature whorl along with the discovery
of some filaments pointing towards the South-West \citep{Sanders05}
for the Southern Ancient bubble all point towards the fact that these
features are the remnants of earlier generations of bubbles.

What can be seen from the phase angles at which the bubbles detach is
that they do not necessarily detach at the same time.  It seems
likely that the jets produced by the SMBH are of the same composition
and power, and as such provide the same amount of energy at the same
rate into the bubbles.  From this, in the case that the ICM were
perfectly uniform, the bubbles would be expected to grow at very
similar rates and so also detach at very similar times.  However it is
very likely that the ICM is not perfectly uniform, the bubbles
may expand at slightly different rates, detaching at slightly different times.

The change in the phase of the precession between the detaching of the
Ghost and Outer bubbles is very similar (see Table \ref{PrecTimes}), even with an extra
precession cycle.  However
there is a difference of up to $60^{\circ}$ between the detaching of
the Northern and Southern bubbles for a given ``generation'', corresponding to around $5\myr$, a large fraction of
the bubble formation time.  In the case of the change between
the Ancient and Outer bubbles, there is more than a complete
precession cycle difference between the Southern and Northern
bubbles (see Table \ref{CycleNo}).  If the Ghost bubbles detached at similar times, then the
Southern Ancient bubble detached about one precession cycle before the
Northern Ancient bubble.  Whether the Southern Ancient bubble belongs
to a generation older than that of the Northern Ancient bubble, and
the true Southern Ancient bubble has not yet been identified, or if
the bubbles did detach at vastly different times is currently unclear.  

This implies that, if the precession is long lived, there are cycles
in the precession when no bubbles detach.  This may mean that the SMBH
is dormant and its duty cycle is less than 100 per cent.  What may
also happen is that the jet does not have sufficient power to inflate
a bubble, but is still supplying energy to the central regions of the
cluster by other mechanisms.

If this were the case, then at times when the jet were less powerful,
and possibly less energy were being supplied to the centre the ICM may
cool and flow inwards for a while.  When the jets become more
powerful, they interact with the cool gas and lift it up out of the
centre, the result of which we see as the whorl.

The spirals seen in the optical emission from NGC1275 \citep{Holtzman92, Carlson98}, which are in
the same sense as the low-temperature whorl are explained if we are
seeing them from below.  The precession of the jet is anti-clockwise
when seen from above, however the precession axis is tilted away from
the line of sight, and so the end of a constant length precessing jet
would appear to move in a clockwise direction, which is the same sense
as the spirals seen in NGC1275.  Their inclination means that we see
them apparently move clockwise.

The best fitting parameters for the precession of 3C84 are consistent
with those presented in \citet{Gower_1982} who compared models of
precessing jets to the observed radio emission.  Their largest timescale
is $\sim10^7\yr$ which is close to the one obtained here.  The opening
angles of the precession have a large range, with an average of around
$\sim30^{\circ}$ which is smaller than the one found for 3C84, but not
inconsistent.

Recent work by \citet{LodatoPringle} show that if black hole's and the accretion
disc's angular momenta were counter-aligned to start with, then over the
course of what would be one precession period the spins would align
(in the case where the disc angular momentum is large compared to the
hole's).  During the course of this alignment the spin axis of the
black hole traces out an arc on the sky.  The resultant position 
angles of the radio jets, and hence bubbles, could give the appearance
of steady precession in the case of the Perseus Cluster.

\section{What could cause the Precession?}\label{causeprec}

Although a precessing jet model is a possible explanation for the
observation that the bubbles occur at different position angles in the
cluster, what is not clear is why the black hole would be precessing.
In a recent comparison between some bipolar Planetary Nebulae and
the bubbles observed in clusters, there are two possible mechanisms
causing the precession \citep{PizzolatoSoker}.  Firstly there could be
a binary black hole at the centre of the NGC1275 galaxy, causing the
disc of the primary to precess \citep{Katz97}.  The second suggestion is that
there is an instability in the disc which warps it and so causes the
precession \citep{Pringle97}.  The timescales predicted from these two models are
similar; for the warped disc model
\[
\tau_{\rm prec} \simeq 2 \times 10^7 \yr\quad
\alpha^{-1}\frac{M}{10^9 \mathcal M_{\odot}},
\]
(Equation 4.11 in \citet{Pringle97}) where $\alpha$ is the
\citet{Shakura_Sunyaev73} viscosity alpha-parameter; and for the binary model
\[
\tau_{\rm prec} \simeq 10^6\yr\Big(\frac{M}{10^9 \mathcal
  M_{\odot}}\Big)^{\frac{1}{2}}\Big(\frac{a}{10^{19}{\rm cm}}\Big)^3\Big(\frac{a_d}{10^{18}{\rm
    cm}}\Big)^{-\frac{3}{2}}\frac{(1+q)^{\frac{1}{2}}}{q \cos \vartheta},
\]
where $q=M_2/M_1$, $M_1$ is the mass of the accreting black hole,
$M_2$ is the mass of the object loosing mass, and $M=M_1+M_2$ is the
total mass of the binary system.  The separation between the
components is $a$ for a circular orbit, $a_d$ is the disc radius and
$\vartheta$ is the tilt angle between the orbital plane and the disc
angle.  \citet{PizzolatoSoker} prefer the binary-driven precession
model to the disc instability driven one because there are point
symmetries in the bubbles in the clusters, assumed to be caused by a
precessing jet.  The warped-disc model would cause a stochastic rather
than a regular precession.  Both scenarios could produce precession times close to those inferred from the bubble directions.

\section{Bubble Dynamics}\label{daVinci}

In a still fluid, small gas bubbles rise rectilinearly, however larger
ones follow zig-zag or spiral paths.  This phenomenon has been called
``Leonardo's Paradox'' as the first recorded observations and attempts at
explanations of this behaviour were performed by Leonardo da Vinci
\citep{Prosperetti04, LeonardodV}.  Since Leonardo's time there have been many attempts to
explain the phenomenon and determine the parameters which cause the
change between linear and periodic motion.

\begin{centering}
\begin{table} \caption{\scshape \label{fluiddyn} Fluid Dynamics}
\begin{tabular}{ll@{ $=$ }l}
Reynolds Number & $\mathcal{R}e$&$2Ur/\nu$ \\
Weber Number$^1$ & $\mathcal{W}e$&$U^2\rho r/\sigma$\\
Galileo Number &
$\mathcal{G}$&$(\sqrt{|\rho_0/\rho-1|g(2r)^3})/\nu$ \\
Strouhal Number & $\mathcal{S}t$&$fr/\langle U_\infty \rangle$ \\
Froude Number & $\mathcal{F}r$&$U/\sqrt{gr}$\\
Drag Coefficient & $C_D$&$4\mathcal{G}^2/3\mathcal{R}e_\infty^2$\\
Fluid Parameter&$\mathcal{M}$&$g\mu^4/\rho\sigma^3 $\\
\end{tabular}
\begin{quote}
$U$($\langle U_\infty \rangle$) is the (average) bubble velocity, $r$ the bubble radius$^2$, $\nu$ is the kinematic viscosity
($\mu/\rho$),  $\rho$ and $\rho_0$ is the density of the fluid and
sphere respectively,  $\sigma$ is the surface tension, $f$ is the
frequency of oscillation of the bubble motion, $\mathcal{R}e_\infty$
is the asymptotic Reynolds number, $g$ is the gravitational
acceleration
$^1$Our definition of $\mathcal{W}e$ is the square of that in
\citet{HartunianSears}.
$^2$In some cases an equivalent radius
$r_{\rm e}=(3\times{\rm Volume}/4\pi)^{1/3}$ is used instead.
\end{quote}
\end{table}
\end{centering}

The motion of spheres and bubbles in fluids has been studied
experimentally as well as with numerical simulations.  Using solid spheres may
not at first sight seem appropriate, but if the liquid is not pure
(``contaminated'') then surfactants on the fluid-bubble interface
(partly) immobilise the fluid, similar to the conditions found on the
surface of a solid object.  In this section we review some of the work on bubbles rising
in fluids to see whether any of these effects could arise for the
bubbles in the Perseus Cluster and cause the bubbles to be found at
different position angles.  In the course of these studies the
fluids in which the bubbles are moving and the motions of the bubbles
have been characterised by dimensionless quantities which are shown in
Table \ref{fluiddyn}.  

The fluid in which the bubble
moves is described by a dimensionless $\mathcal{M}$.   The behaviour of  $C_{\rm
  D}$ with $\mathcal{R}e$ depends on the $\mathcal{M}$ of the fluid.
Fluids with $10^{-10}<\mathcal{M}<10^{-8}$ (e.g. filtered water, methanol) are ``fast''
and show the characteristic dip in $C_{\rm D}$ at
$100<\mathcal{R}e<1000$, which corresponds to $\mathcal{W}e\sim
1.5$.  ``Slow'' fluids (e.g. mineral oil, water-glycerin mix) have 
$10^{-7}<\mathcal{M}<10^{-2}$ and do not show this dip
\citep{HartunianSears}.

The critical threshold for the onset of the non-rectilinear motion depends
on the purity of the liquid -- the Reynolds number being important for
contaminated liquids ($\mathcal{R}e_{\rm crit}\sim 210$) and the
Weber number for pure liquids ($\mathcal{W}e_{\rm crit}\sim 1.58$).  

Numerical simulations show that the wake
behind a sphere which is held {\it fixed} in a uniform flow is
axisymmetric up to $\mathcal{R}e=212$. Above this value a planar symmetric wake is found, consisting of two
steady counter-rotating threads.  At $\mathcal{R}e\approx270$ the
planar symmetric wake become time-dependent; oppositely signed
stream-wise vortices form a series of loops resembling hairpin
vortices.  Eventually, as $\mathcal{R}e$ continues to increase the wake behind the
sphere becomes turbulent.  The exact values of $\mathcal{R}e$ depend
on the density ratio, $\rho_0/\rho$ \citep{Jenny04}.  The results do not change dramatically if
the sphere is moving freely in a still fluid.  \citet{deVries} investigate bubble motion as opposed to sphere motion
and state that in the case of bubbles $\mathcal{R}e_{\rm crit}\sim 600$, however this discrepancy may arise from the purity of the
fluid.

Using numerical simulations, \citet{Jenny04} investigated the instabilities in the motions of
spheres (rising and falling) in Newtonian fluids finding that $\mathcal{G}_{\rm
  crit}\sim170$, for a density ratio of $\rho_0/\rho=0$, with a
$\mathcal{S}t=0.05$.  They find that a range of behaviours occur for bubbles moving with
different density ratios and Galileo number (see their Fig. 29).  For a
given $\rho_0/\rho$, the wake behind a
sphere starts axisymmetric, then as the Galileo number increases
the motion becomes steady but oblique.
Continuing to increase $\mathcal{G}$ causes oscillations to occur
during the oblique ascent, eventually causing zig-zag motion.
Eventually at $\mathcal{G}\sim 200$ the motion becomes chaotic.

There are two types of motion when the wake becomes time dependent --
helical and zig-zag.  The transition between the two types of motion
is thought to depend on the bubble size, observed experimentally by
\citet{LundePerkins97}.  Larger bubbles deform from spherical shapes
to ellipsoidal ones and numerical simulations by
\citet{MouginMagnaudet} show that the transition between the two types
of periodic motion occurs for an aspect
ratio, $\chi=2.5$.  

\citet{HartunianSears} state that the path instability does not
occur for small and spherical-cap bubbles, but only for
intermediate sized ellipsoidal bubbles, where the motion ranges from
oscillations to gentle rocking about the direction of motion.  It is
possible for the zig-zag motions to cause deflections from a vertical
path by to $50^{\circ}$ \citep{LundePerkins97} which is similar to the
position angles of the North-Western Ghost and the South-Western
Ancient bubbles.

\subsection{Spherical Cap Bubbles}\label{spherical}

\citet{WegenerParlange} show that $\mathcal{F}r=1.00\pm0.05$ for
spherical cap bubbles.  In the
assumption that this relation continues at much larger volumes, then for the
North-western Ghost bubble, the buoyancy velocity calculated during
the work of \citet{DunnFabian05} is $\sim3.6\times10^7{\rm\thinspace
  cm~s^{-1}}$ and the radius of the bubble is $\sim14 \kpc$.  Using the
enclosed mass and $R_{\rm dist}$ to obtain an estimate on the gravitational field, $g$,
gives $\mathcal{F}r\approx0.92$.  Therefore, if the relation in
\citet{WegenerParlange} holds up to these bubble sizes, then it is
likely that the observed X-ray decrement is, and behaves like, a
spherical cap bubble.

In this case it could be assumed to follow the behaviour observed by
\citet{HartunianSears}, i.e. it rises rectilinearly.  Therefore it
could be assumed that the other Ghost and Ancient bubbles would have
travelled linearly out from the cluster after they had attained the
spherical-cap shape.

\subsection{Cluster Bubbles}

The bubbles in the Perseus Cluster as analysed by
\citet{ACF_Halpha_PER03} inferred that the ICM was
viscous, and as such the Reynolds number would be expected to be less
than $1000$.  The kinematic viscosity inferred from this Reynolds number and the
North-western Ghost bubble is $4\times 10^{27} \cmsqps$.  Therefore it
could reasonably be
assumed that the ICM is a ``slow, contaminated'' fluid and as such the Reynolds
number rather than the Weber number would determine the onset of path
instability in the bubbles.

\citet{WegenerParlange} state that spherical cap bubbles rise steadily
with no periodic motion, and so the path instability would apply to
the newly formed bubbles as they would be spherical or ellipsoidal.  The
 critical Reynolds number for the onset of the path
instability is $\sim200$.  Therefore, as the upper limit on the
Reynolds number of the ICM is $\sim 1000$, the bubbles could have been
oscillating about their direction of motion shortly after their
creation.  As the bubbles' shape evolved into a spherical cap their
oscillations reduced and they may then end up travelling at a
different position angle than that of the jet.

For the South-western Outer bubble, using the kinematic viscosity from
\citet{ACF_Halpha_PER03}, and assuming
that the density ratio $\rho_0/\rho\sim0$, then the Galileo number is
$\sim 1140$, which is in the regime of chaotic motion.  This is
unlikely as there are indicators in the X-ray and H$\alpha$ emission
that the motion of the bubbles has been fairly steady.  As such the Galileo number would be expected to be much
lower, $\lesssim200$.  This implies that the Reynolds number, which has been
used to estimate the viscosity, would be around $200$, giving a
viscosity of $2\times10^{28}\cmsqps$.  There is no large change in these values if
$\rho_0/\rho<1$ were used instead.

The mechanism which causes the onset of path instability in the rising
bubbles is currently uncertain.  However, if the onset of the path
instability occurs at the same values of the Galileo and Reynolds
numbers at the scales observed in clusters of galaxies, then, as the
calculated values for these numbers fall in the range where the bubble
motion would be expected to be non-rectilinear, the bubbles could
oscillate as they rise.  This motion, however, would not last for long
as the bubbles appear to evolve into spherical caps (which show no
oscillatory motion, see Section \ref{spherical}) comparatively quickly, before complete
oscillations would have taken place.  The position angles of the inner bubbles are such that the deviations
from a linear rise would have to start very quickly, almost
immediately after the bubbles detached.  

The results of the experiments discussed in the above sections have
all been performed on the Earth where the gravitational potential can
be regarded as 1-dimensional.  In the environment of the cluster the
gravitational potential is spherically symmetric.  What effect this
would have on bubbles which are oscillating about their direction of
motion is not clear, and probably depends on relative sizes of the
bubble to the cluster potential.  Bubbles moving obliquely in experiments on
Earth may move in spirals in a cluster, which could be an explanation for
the low temperature structure observed in the temperature maps of
\citet{SandersMap04} if the bubbles did not evolve into a spherical-cap shape.  

However this does depend on there being sufficient time
for the oscillations to occur before the bubble evolves into a
spherical cap shape; the Southern Outer bubbles' shape as seen in the X-rays
already appears similar to that of a spherical cap.  If the Strouhal
number relation also holds up to these bubble volumes, then
$f\approx10^{-16}{\rm \thinspace s^{-1}}\approx 3 \gyr^{-1}$.  This
implies that there would be no chance for any oscillation before the
bubble shape evolves into a spherical cap. 

The radial arrangement of the H$\alpha$ filaments in the cluster strongly
suggest that the bubbles move in a linear fashion, as oscillating bubbles, or bubbles
moving in spirals would be expected to drag the filaments with them as
they rise.

\section{Conclusions}

The observations of the bubbles in the Perseus Cluster show that they
are at different position angles with respect to the cluster
centre.  We have outlined several interpretations for this behaviour,
including a jet which is precessing or moving around.  For the
precessing jet we find
that the best fit precession parameters are that the projected
precession angle offset is $10^{\circ}$ westwards of North; the precession axis is
tilted $120^{\circ}$ away from the line of sight and the precession
opening angle is $50^{\circ}$.  Using the timescales for the bubbles
the precession timescale is around $3.3 \times 10^7 \yr$.  We review some work on the motion on bubbles rising in fluids and the
onset of path instability.  The conditions of the motion in the
cluster are such that the oscillations about the direction of motion
are possible for the young bubbles, but the timescales are of order
$1\gyr$, by which time the bubbles have evolved into spherical caps,
which do not exhibit this type of motion.  A steady precession or motion of the jet may be responsible for the
whorl seen in the temperature structure. Alternatively, we note that
the whorl may be an artifact of a partial inner circle and a partial
outer ellipse, connected together in the SE. The structures at the
centre of the Perseus cluster provide an enigmatic view of the past
history of this region.

\section*{Acknowledgements}

ACF and RJHD acknowledge support from The Royal Society
and PPARC respectively.

\bibliographystyle{mnras} 
\bibliography{mn-jour,/home/rjhd2/bibtex/dunn}

\begin{thebibliography}{}

\bibitem[\protect\citeauthoryear{{Barker} \& {Byrd}}{{Barker} \&
  {Byrd}}{1981}]{Barker1981}
{Barker} B.~M.,  {Byrd} G.~G., 1981, \apjl, 245, L67

\bibitem[\protect\citeauthoryear{{Blundell} \& {Bowler}}{{Blundell} \&
  {Bowler}}{2005}]{Blundell}
{Blundell} K.~M.,  {Bowler} M.~G., 2005, \apjl, 622, L129

\bibitem[\protect\citeauthoryear{{B{\"o}hringer} et~al.}{{B{\"o}hringer}
  et~al.}{1993}]{Bohringer}
{B{\"o}hringer} H., {Voges} W., {Fabian} A.~C., {Edge} A.~C.,  {Neumann} D.~M.,
  1993, \mnras, 264, L25

\bibitem[\protect\citeauthoryear{{Bridle} et~al.}{{Bridle}
  et~al.}{1979}]{Bridle1979}
{Bridle} A.~H., {Davis} M.~M., {Fomalont} E.~B., {Willis} A.~G.,  {Strom}
  R.~G., 1979, \apjl, 228, L9

\bibitem[\protect\citeauthoryear{{Carlson} et~al.}{{Carlson}
  et~al.}{1998}]{Carlson98}
{Carlson} M.~N. et~al., 1998, \aj, 115, 1778

\bibitem[\protect\citeauthoryear{{Churazov} et~al.}{{Churazov}
  et~al.}{2000}]{Churazov00}
{Churazov} E., {Forman} W., {Jones} C.,  {B{\" o}hringer} H., 2000, \aap, 356,
  788

\bibitem[\protect\citeauthoryear{{Conselice}, {Gallagher}, \&
  {Wyse}}{{Conselice} et~al.}{2001}]{Conselice01}
{Conselice} C.~J., {Gallagher} J.~S.,  {Wyse} R.~F.~G., 2001, \aj, 122, 2281

\bibitem[\protect\citeauthoryear{{da Vinci}}{{da Vinci}}{1515}]{LeonardodV}
{da Vinci} L., {c. 1505-1515}, {\it Codex Leicester}, {E. MacCurdy, {\it The
  Notebooks of Leonardo da Vinci} (Reynal \& Hitchcock, New York, 1938) Vols. 1
  \& 2}

\bibitem[\protect\citeauthoryear{{de Vries}, {Biesheuvel}, \& {van
  Wijngaarden}}{{de Vries} et~al.}{2002}]{deVries}
{de Vries} A.~W.~G., {Biesheuvel} A.,  {van Wijngaarden} L., 2002, Int. J. of
  Multiphase Flow, 28, 1823

\bibitem[\protect\citeauthoryear{{Dunn} \& {Fabian}}{{Dunn} \&
  {Fabian}}{2004}]{DunnFabian04}
{Dunn} R.~J.~H.,  {Fabian} A.~C., 2004, \mnras, 351, 862

\bibitem[\protect\citeauthoryear{{Dunn}, {Fabian}, \& {Taylor}}{{Dunn}
  et~al.}{2005}]{DunnFabian05}
{Dunn} R.~J.~H., {Fabian} A.~C.,  {Taylor} G.~B., 2005, \mnras, accepted,
  astro-ph/0510191

\bibitem[\protect\citeauthoryear{{Ekers} et~al.}{{Ekers}
  et~al.}{1978}]{Ekers1978}
{Ekers} R.~D., {Fanti} R., {Lari} C.,  {Parma} P., 1978, \nat, 276, 588

\bibitem[\protect\citeauthoryear{{Fabian} et~al.}{{Fabian}
  et~al.}{2002}]{Celotti02}
{Fabian} A.~C., {Celotti} A., {Blundell} K.~M., {Kassim} N.~E.,  {Perley}
  R.~A., 2002, \mnras, 331, 369

\bibitem[\protect\citeauthoryear{{Fabian} et~al.}{{Fabian}
  et~al.}{2003a}]{ACF_deep_PER03}
{Fabian} A.~C., {Sanders} J.~S., {Allen} S.~W., {Crawford} C.~S., {Iwasawa} K.,
  {Johnstone} R.~M., {Schmidt} R.~W.,  {Taylor} G.~B., 2003a, \mnras, 344, L43

\bibitem[\protect\citeauthoryear{{Fabian} et~al.}{{Fabian}
  et~al.}{2003b}]{ACF_Halpha_PER03}
{Fabian} A.~C., {Sanders} J.~S., {Crawford} C.~S., {Conselice} C.~J.,
  {Gallagher} J.~S.,  {Wyse} R.~F.~G., 2003b, \mnras, 344, L48

\bibitem[\protect\citeauthoryear{{Fabian} et~al.}{{Fabian}
  et~al.}{2003c}]{ACF_3C294}
{Fabian} A.~C., {Sanders} J.~S., {Crawford} C.~S.,  {Ettori} S., 2003c, \mnras,
  341, 729

\bibitem[\protect\citeauthoryear{{Fabian} et~al.}{{Fabian}
  et~al.}{2000}]{ACF_complex_PER00}
{Fabian} A.~C. et~al., 2000, \mnras, 318, L65

\bibitem[\protect\citeauthoryear{{Gower} et~al.}{{Gower}
  et~al.}{1982}]{Gower_1982}
{Gower} A.~C., {Gregory} P.~C., {Unruh} W.~G.,  {Hutchings} J.~B., 1982, \apj,
  262, 478

\bibitem[\protect\citeauthoryear{{Gull} \& {Northover}}{{Gull} \&
  {Northover}}{1973}]{GullNorthover}
{Gull} S.~F.,  {Northover} K.~J.~E., 1973, \nat, 244, 80

\bibitem[\protect\citeauthoryear{{Hartunian} \& {Sears}}{{Hartunian} \&
  {Sears}}{1957}]{HartunianSears}
{Hartunian} R.~A.,  {Sears} W.~R., 1957, J. Fluid Mech., 3, 27

\bibitem[\protect\citeauthoryear{{Hatch} et~al.}{{Hatch}
  et~al.}{2005}]{Hatch2005inprep}
{Hatch} N.~A., {Crawford} C.~S., {Fabian} A.~C.,  {Johnstone} R.~M., 2005, in
  prep.

\bibitem[\protect\citeauthoryear{{Hjellming} \& {Johnston}}{{Hjellming} \&
  {Johnston}}{1981}]{HjellmingJohnston81}
{Hjellming} R.~M.,  {Johnston} K.~J., 1981, \apjl, 246, L141

\bibitem[\protect\citeauthoryear{{Holtzman} et~al.}{{Holtzman}
  et~al.}{1992}]{Holtzman92}
{Holtzman} J.~A. et~al., 1992, \aj, 103, 691

\bibitem[\protect\citeauthoryear{{Jenny}, {Du\v{s}ek}, \& {Bouchet}}{{Jenny}
  et~al.}{2004}]{Jenny04}
{Jenny} M., {Du\v{s}ek} J.,  {Bouchet} G., 2004, J. Fluid Mech., 508, 201

\bibitem[\protect\citeauthoryear{{Katz}}{{Katz}}{1997}]{Katz97}
{Katz} J.~I., 1997, \apj, 478, 527

\bibitem[\protect\citeauthoryear{{Krichbaum} et~al.}{{Krichbaum}
  et~al.}{1992}]{Krichbaum92_Per}
{Krichbaum} T.~P. et~al., 1992, \aap, 260, 33

\bibitem[\protect\citeauthoryear{{Lodato} \& {Pringle}}{{Lodato} \&
  {Pringle}}{2005}]{LodatoPringle}
{Lodato} G.,  {Pringle} J.~E., 2005, \mnras, submitted

\bibitem[\protect\citeauthoryear{{Lunde} \& {Perkins}}{{Lunde} \&
  {Perkins}}{1997}]{LundePerkins97}
{Lunde} K.,  {Perkins} R.~J., 1997, ASME Fluids Engineering Division Summer
  Meeting, FEDSM97

\bibitem[\protect\citeauthoryear{{Lynds}}{{Lynds}}{1970}]{Lynds70}
{Lynds} R., 1970, \apjl, 159, L151

\bibitem[\protect\citeauthoryear{{Margon}}{{Margon}}{1984}]{Margon84}
{Margon} B., 1984, \araa, 22, 507

\bibitem[\protect\citeauthoryear{{Minkowski}}{{Minkowski}}{1957}]{Minkowski57}
{Minkowski} R., 1957, in IAU Symp. 4: Radio astronomy, p. 107

\bibitem[\protect\citeauthoryear{{Mougin} \& {Magnaudet}}{{Mougin} \&
  {Magnaudet}}{2002}]{MouginMagnaudet}
{Mougin} G.,  {Magnaudet} J., 2002, Phys. Rev. Lett., 88, 014502

\bibitem[\protect\citeauthoryear{{Pedlar} et~al.}{{Pedlar}
  et~al.}{1990}]{Pedlar90}
{Pedlar} A., {Ghataure} H.~S., {Davies} R.~D., {Harrison} B.~A., {Perley} R.,
  {Crane} P.~C.,  {Unger} S.~W., 1990, \mnras, 246, 477

\bibitem[\protect\citeauthoryear{{Pizzolato} \& {Soker}}{{Pizzolato} \&
  {Soker}}{2005}]{PizzolatoSoker}
{Pizzolato} F.,  {Soker} N., 2005, Elsevier Science, astro-ph/0501658

\bibitem[\protect\citeauthoryear{{Pringle}}{{Pringle}}{1997}]{Pringle97}
{Pringle} J., 1997, \mnras, 292, 136

\bibitem[\protect\citeauthoryear{{Prosperetti}}{{Prosperetti}}{2004}]{Prospere%
tti04}
{Prosperetti} A., 2004, Phsyics of Fluids, 16, 1852

\bibitem[\protect\citeauthoryear{{Sanders} et~al.}{{Sanders}
  et~al.}{2004}]{SandersMap04}
{Sanders} J.~S., {Fabian} A.~C., {Allen} S.~W.,  {Schmidt} R.~W., 2004, \mnras,
  349, 952

\bibitem[\protect\citeauthoryear{{Sanders}, {Fabian}, \& {Dunn}}{{Sanders}
  et~al.}{2005}]{Sanders05}
{Sanders} J.~S., {Fabian} A.~C.,  {Dunn} R.~J.~H., 2005, \mnras, 360, 133

\bibitem[\protect\citeauthoryear{{Shakura} \& {Sunyaev}}{{Shakura} \&
  {Sunyaev}}{1973}]{Shakura_Sunyaev73}
{Shakura} N.~I.,  {Sunyaev} R.~A., 1973, \aap, 24, 337

\bibitem[\protect\citeauthoryear{{Walker} et~al.}{{Walker}
  et~al.}{2000}]{Walker2000}
{Walker} R.~C., {Dhawan} V., {Romney} J.~D., {Kellermann} K.~I.,  {Vermeulen}
  R.~C., 2000, \apj, 530, 233

\bibitem[\protect\citeauthoryear{{Wegener} \& {Parlange}}{{Wegener} \&
  {Parlange}}{1973}]{WegenerParlange}
{Wegener} P.~P.,  {Parlange} J.-Y., 1973, Ann. Rev. Fluid Mech., 5, 79

\end{thebibliography}

\end{document}